\documentclass[letterpaper,pre,showpacs,showkeys,english,preprint,aps]{revtex4-1}
\usepackage[T1]{fontenc}
\usepackage[latin9]{inputenc}
\usepackage{mathrsfs}
\usepackage{amsmath}
\usepackage{amssymb}
\usepackage{graphicx}
\usepackage{esint}
\PassOptionsToPackage{normalem}{ulem}
\usepackage{ulem}



\usepackage{babel}
\begin{document}

\title{A path integral formalism for the closure\\
of autonomous statistical systems}

\author{Richard Kleeman}

\affiliation{{\small{}Department of Mathematics, Courant Institute of Mathematical
Sciences, New York, USA.}}

\date{\today}
\begin{abstract}
Recently a path integral formalism has been proposed by the author
which gives the time evolution of moments of slow variables in a Hamiltonian
statistical system. This closure relies on evaluating the informational
discrepancy of a time sequence (path) of approximating densities from
the Liouvillian evolution that an exact density must follow. The discrepancy
is then used to weight all possible paths using a generalized Boltzmann
principle. That formalism is extended here to deal with more general
and realistic autonomous dynamical systems. There the divergence of
the time derivative of dynamical variables need not vanish as it does
in the Hamiltonian case and this property complicates the closure
derivation. Many interesting and realistic applications are covered
by this new formalism including those describing realistic turbulence
and the relevant specifics of this situation are outlined. The practical
issues associated with the implementation of the outlined formalism
are also discussed. 
\end{abstract}

\pacs{05}


\maketitle

\section{Introduction}

The statistical closure of the dynamical systems underlying turbulent
phenomenon is a long standing and practically important problem in
mathematical physics. Recently the author has proposed a new formalism
to deal with Hamiltonian systems using a Wiener path integral \citep{Kle14}.
This formalism builds on earlier work by Turkington \citep{turkington2012optimization}
which has been validated numerically for first moments using an inviscid
truncated Burgers system \citep{kleeman2012nonequilibrium}. The long
term aim of this work is to apply the closure methodology to realistic
turbulent system of various kinds. Consequently it would be desirable
if the Hamiltonian restriction could be relaxed in order to deal with
systems with dissipation and external forcing.

As is well known (see e.g. \citep{salmon-book} Chapter 7) a large
number of important \uline{conservative} fluid dynamical systems
can be cast into Hamiltonian form. Depending on the manner in which
this is done and the particular fluid system, these may be either
canonical or non-canonical. Specifically this means that as dynamical
systems, they may be written as \footnote{For pedagogical reasons we are considering a finite set of dynamical
system variables which implies some kind of truncation of the fluid
system at a fine scale. Generalizations to a continuum of variables
are straighforward. Note also that the summation convention is being
used on repeated indices.} 
\begin{eqnarray*}
\frac{dx_{i}}{dt} & = & \left\{ x_{i},H\right\} =J_{ij}\frac{\partial H}{\partial x_{j}}\equiv A_{i}(x)\\
\left\{ Q,P\right\}  & = & J_{ij}\frac{\partial P}{\partial x_{j}}\frac{\partial Q}{\partial x_{i}}
\end{eqnarray*}
where $H$ is the Hamiltonian and the Poisson bracket $\left\{ ,\right\} $
is antisymmetric and satisfies the (Lie) Jacobi identity. Due to the
bracket properties, the covariant tensor $J_{ij}$ is anti-symmetric
\footnote{Note that $J$ transforms as a second rank tensor under non-singular
changes of dynamical variables.}. Furthermore an appropriate non-singular transformation of dynamical
variables allows it to be brought into the form of a constant anti-symmetric
matrix (see \citep{salmon-book} p333-337). For such a set of variables
it is trivial to confirm that
\begin{equation}
\frac{\partial A_{i}}{\partial x_{i}}=0\label{divergencefree}
\end{equation}
and since $A_{i}$ is a covariant vector, this non-divergent condition
holds for any non-degenerate choice of variables. As is also very
well known such a condition is crucial for establishing that the statistical
density evolution equation for an ensemble of dynamical system trajectories
is the Liouville equation:
\begin{equation}
\varrho_{t}+A_{i}\frac{\partial}{\partial x_{i}}\varrho=0\label{classical}
\end{equation}

This equation implies easily that any function of the density also
satisfies the same equation. Furthermore condition (\ref{divergencefree})
implies that the Liouvillian operator $L\equiv A_{i}\frac{\partial}{\partial x_{i}}$
is anti-Hermitean with respect to the real $L^{2}$ inner product
(see below for more detail on the more general case). Both of these
properties were needed to derive the closure in \citep{Kle14} and
so the non-divergence condition (\ref{divergencefree}) was obviously
essential. 

Consider now a conventional turbulent fluid system. As noted in \citep{salmon-book}
p233, this may often be described by the following dynamical system
\begin{equation}
\frac{dx_{i}}{dt}=A_{i}(x)+\alpha(i)x_{i}+F(i)\label{Generic}
\end{equation}
where brackets indicate no summation convention and where the $A_{i}$
belong to a conservative system which usually can be brought into
Hamiltonian form as discussed above. The second term on the right
denotes a (linear) dissipation (i.e. $\alpha\leq0)$ while the third
denotes a possible external and prescribed injection of energy. Such
a system is often called a forced dissipative model of turbulence.
If $F$ is absent we obtain a decaying turbulence model. The differences
of these two situations in the context of the present closure are
discussed further below. The Navier Stokes equation, appropriately
truncated spectrally, may be cast into the form of (\ref{Generic})
with the coefficients $\alpha(i)$ increasing strongly with increasing
wavenumber $i$. Physically the dissipation there is intended to represent
molecular diffusion. Other turbulent systems such as a quasi-geostrophic
fluid may have quite different forms for both the dissipation and
$A_{i}$. In anycase the generic form of the dynamical system equations
implies that
\[
\frac{\partial A_{i}}{\partial x_{i}}=\sum_{i}\alpha(i)<0
\]

which must be true no matter what dynamical variables are chosen to
describe the system. As discussed in detail below this non-zero divergence
means that a non-trivial function of the statistical density no longer
satisfies a generalized Liouville equation and also that the generalized
Liouvillian operator is no longer anti-Hermitean. We therefore extend
our closure derivation to an autonomous rather than a Hamiltonian
dynamical system to deal with this more realistic situation. Our closure
will be seen to be appropriate for a completely general autonomous
dynamical system not just that of (\ref{Generic}) and so further
applications beyond conventional turbulence systems will also be possible.

In section 2 we briefly review the earlier work mainly from a motivational
and scene setting perspective before providing a complete derivation
of our formalism for a general autonomous dynamical system in Sections
3 and 4. A discussion of applications may be found in Section 5. The
reader is referred to the earlier papers mentioned above for more
details on the Hamiltonian system closure methodology.

\section{Conceptual description of the closure formalism}

\subsection{Onsager-Machlup Theory}

It has been well known since early in the history of statistical physics
that the probability of a fluctuation from an equibrium state at a
fixed time is given by the Boltzmann principle
\[
p(\lambda)=C\exp\left(-\sigma\left(\lambda\right)\right)
\]

where $\sigma$ is the \textbf{entropy} function for the system. Onsager
and Machlup (OM) \citep{onsager1953fluctuations} generalized this
principle in the early 1950s and defined a positive likelihood weight
$W$ on temporal \textbf{paths} $\lambda(t)$ of near equilibrium
states
\begin{equation}
W\left[\lambda(t)\right]=\exp\left(-\int_{0}^{t}\mathscr{L}\left[\lambda(t)\right]dt\right)\label{path weight}
\end{equation}

The ``Lagrangian'' $\mathscr{L}$ was assumed to be of the form:
\[
\mathscr{L}\left[\lambda\left(t\right)\right]=\frac{1}{2}\left(\dot{\lambda}-U\mathbf{\lambda}\right)^{t}g\left(\dot{\lambda}-U\mathbf{\lambda}\right)
\]

where $U$ and $g$ are constant matrices with the latter importantly
assumed to be positive definite. 

It is immediately clear from the latter assumption that the path of
greatest weight is simply given by
\begin{equation}
\dot{\lambda}=U\mathbf{\lambda}\label{drift}
\end{equation}

which is, in general, a relaxation to equilibrium for appropriate
choices for $U$. Furthermore the probability of a particular $\lambda$
at time $T$ is obtained by \textbf{integrating} over the likelihood
weights of all paths ending at $\lambda(T)$:
\begin{equation}
p(\lambda\left(T\right))=C\int D\lambda W\left[\lambda\right]\label{pathintegral}
\end{equation}

which is a functional or \textbf{path integral} over all possible
paths which end at $\lambda\left(T\right)$. As is well known \citep{risk89}
this situation corresponds with an Ornstein Uhlenbeck stochastic process
and the most likely $\lambda$ also follows equation (\ref{drift}).
Such a trajectory is called thermodynamical since it is most probable.

Many attempts have been made to formulate far from equilibrium versions
of OM theory. The approach here follows three principles:
\begin{enumerate}
\item The Lagrangian $\mathscr{L}$ should be derivable from a fundamental
information theoretic argument relating to paths since it effectively
generalizes the entropy function underlying the original Boltzmann
principle.
\item The precise form of the Lagrangian should be deducible from first
principles and reflect the statistical properties of the slow or coarse
grained variables for the statistical system.
\item In appropriate limits thermodynamical relations of the type proposed
recently by Öttinger \citep{oett} should be recovered since such
theories work well in practical applications.
\end{enumerate}

\subsection{Trial density manifold}

Zubarev \citep{Zub74} proposed the concept of a trial density as
a way of approximating densities of non-equilibrium states. One specifies
a set of moments of slow variables as thermodynamical variables and
then uses a maximum entropy principle to associate these with a ``trial''
density. In general such trial densities $\hat{\varrho}$ are not
exact densities $\varrho$ for the given non-equilibrium state with
the given moments. This situation contrasts with the equilibrium Gibbs
density. The philisophical approach taken here therefore differs from
that adopted in equilibrium statistical physics in that it is tacitly
assumed that the \textbf{exact} density for a system will \textbf{never
generally} be available and thus one must deal always with an approximation
manifold of densities. On such a manifold the ``veracity'' of each
density will be assigned a non-negative weight which we shall term
a \textbf{consistency distribution}. It is analogous to a quantum
wavefunction as we shall see below.

The set of all trial densities with a prescribed set of moments can
be given a Riemannian manifold structure by using the associated Fisher
information matrix as a metric tensor. This manifold structure is
the essence of the subject of information geometry \citep{ama00}
and shall become important to our discussion later.

\section{Application to an autonomous dynamical system}

\subsection{Liouvillian discrepancy}

The equation for the density evolution within a general autonomous
dynamical equation is simply the Fokker-Planck equation without a
diffusion term \footnote{We refer to this equation as a \uline{generalized} Liouville equation
as opposed to the classical such equation of (\ref{classical}).} i.e.
\begin{eqnarray}
\varrho_{t}+L\varrho & = & 0\label{Liouville}\\
L & \equiv & \frac{\partial}{\partial x_{i}}A_{i}\nonumber 
\end{eqnarray}

with summation convention and where the original dynamical system
is
\[
\frac{dx_{i}}{dt}=A_{i}(x)
\]

As usual the formal adjoint of the operator $L$ is given by
\[
L^{*}=-A_{i}\frac{\partial}{\partial x_{i}}
\]

From this we deduce that
\begin{equation}
L+L^{*}=L^{d}\equiv\frac{\partial A_{i}}{\partial x_{i}}\label{adjoint}
\end{equation}
Consider now a temporal trajectory or path through the trial density
manifold. In general this will not satisfy equation (\ref{Liouville}).
A measure of the discrepancy of a given path from such an evolution
can be defined as follows: Suppose the trial path $\hat{\varrho}(t)$
is evolved forward in time an additional $\Delta t$ according to
the above equation resulting in
\begin{eqnarray*}
\varrho'(t+\Delta t) & = & \exp\left(-\Delta tL\right)\hat{\varrho}(t)=\exp\left(\Delta tT\right)\hat{\varrho}(t)\\
T & \equiv & \frac{\partial}{\partial t}
\end{eqnarray*}

According to elementary information theory the ``distance'' between
this evolved density and the assumed trial density $\hat{\varrho}(t+\Delta t)$
can be measured by their relative entropy i.e. by $D\left(\varrho'(t+\Delta t)||\hat{\varrho}(t+\Delta t)\right)$.

Now we require the time evolution of $l\equiv\log\varrho$ to calculate
the information loss rate defined using the relative entropy. Unlike
the Hamiltionian case for which $\frac{\partial A_{i}}{\partial x_{i}}=0$
arbitrary functions of the density do not satisfy equation (\ref{Liouville})
so we need to compute the evolution of $l$ explictly using a Taylor
series. 

A density evolving according to (\ref{Liouville}) will satisfy
\[
\left(T-L^{*}\right)\varrho=-L^{d}\varrho
\]

Further from the form of $L^{*}$ it follows easily using the chain
rule that
\[
\left(T-L^{*}\right)F(\varrho)=F'\left(T-L^{*}\right)\rho=-F'\frac{\partial A_{i}}{\partial x_{i}}\varrho
\]

and so
\[
\left(T-L^{*}\right)l=-\frac{\partial A_{i}}{\partial x_{i}}
\]

or 
\begin{equation}
Tl=L^{*}l-\frac{\partial A_{i}}{\partial x_{i}}\label{logevolve}
\end{equation}

now assuming that $A$ does not depend on time explicitly (the autonomous
assumption) then we have after iteration
\[
T^{n}l=\left(L^{*}\right)^{n}l-\left(L^{*}\right)^{n-1}\frac{\partial A_{i}}{\partial x_{i}}
\]

and so
\[
l(t+\Delta t)=e^{\Delta tL^{*}}l-\left(\Delta t\frac{\partial A_{i}}{\partial x_{i}}+\frac{\left(\Delta t\right)^{2}}{2}L^{*}\left(\frac{\partial A_{i}}{\partial x_{i}}\right)+\ldots\right)
\]

where we are expanding only to second order since this will be appropriate
below.

\subsection{Evolution of a Liouville residual}

Consider now a trial density $\hat{\varrho}$ which does not evolve
according to equation (\ref{Liouville}) since it is constrained to
lie within the trial density manifold. Denote by angle brackets the
expectation with respect to this density at a particular time. For
a general random variable $F$ we have 
\begin{eqnarray}
\frac{\partial\left\langle F\right\rangle }{\partial t}-\left\langle LF\right\rangle  & = & \left\langle F_{t}\right\rangle +\int F\left(T-L^{*}\right)\hat{\varrho}\nonumber \\
 & = & \left\langle F_{t}\right\rangle +\int F\left(T-L^{*}\right)\hat{l}\nonumber \\
 & = & \left\langle F_{t}\right\rangle +\left\langle FR'\right\rangle \label{Feq}\\
R' & \equiv & \left(T-L^{*}\right)\hat{l}\nonumber 
\end{eqnarray}

where we are using integration by parts to re-express $L$ as the
adjoint operator on the first line and using the chain rule on the
second line. Choosing $F=1$ we obtain
\begin{equation}
\left\langle R'+\frac{\partial A_{i}}{\partial x_{i}}\right\rangle \equiv\left\langle R\right\rangle =0\label{Rdef}
\end{equation}

where $R$ is a generalized Liouville residual since it vanishes if
a density evolves according to (\ref{Liouville}) (see equation (\ref{logevolve})).
We can rewrite (\ref{Feq}) as 
\[
\frac{\partial\left\langle F\right\rangle }{\partial t}-\left\langle LF\right\rangle =\left\langle F_{t}\right\rangle +\left\langle FR\right\rangle -\left\langle F\frac{\partial A_{i}}{\partial x_{i}}\right\rangle 
\]

and now set $F=R$ obtaining
\[
-\left\langle LR\right\rangle =\left\langle TR\right\rangle +\left\langle R^{2}\right\rangle -\left\langle R\frac{\partial A_{i}}{\partial x_{i}}\right\rangle 
\]

so
\[
-\left\langle (L+T)R\right\rangle =\left\langle R^{2}\right\rangle -\left\langle R\frac{\partial A_{i}}{\partial x_{i}}\right\rangle 
\]

or using (\ref{adjoint})
\[
-\left\langle (T-L^{*})R\right\rangle =\left\langle R^{2}\right\rangle 
\]

or using the definition of $R$ and $R'$ as 
\begin{equation}
-\left\langle (T-L^{*})^{2}\hat{l}\right\rangle =\left\langle R^{2}\right\rangle +\left\langle (T-L^{*})\frac{\partial A_{i}}{\partial x_{i}}\right\rangle \label{Expand}
\end{equation}

\subsection{Information loss}

For one timestep $\Delta t$ this can be expressed as (see \citep{Kle14})
\begin{align*}
IL & \equiv D(\varrho(t+\Delta t)||\hat{\varrho}(t+\Delta t))\\
 & =\int\varrho(t+\Delta t)\left(l(t+\Delta t)-\hat{l}(t+\Delta t)\right)
\end{align*}

where $\varrho$ and $l$ are the (generalized) Liouville evolved
density and log density over the time step with the starting density
and log density being those drawn from the trial density manifold
at that earlier time i.e. $\varrho(t)=\hat{\varrho}(t)\quad l(t)=\hat{l}(t)$.
\begin{widetext}
Using (\ref{Liouville}) and the log density counterpart (\ref{logevolve})
this may be re-expressed to second order as
\[
IL=\int e^{-\Delta tL}\hat{\varrho}(t)\left(e^{-\Delta tL^{*}}\hat{l(}t)-\left(\Delta t\frac{\partial A_{i}}{\partial x_{i}}+\frac{\left(\Delta t\right)^{2}}{2}L^{*}\left(\frac{\partial A_{i}}{\partial x_{i}}\right)+\ldots\right)-e^{\Delta tT}\hat{l}(t)\right)
\]

Using integration by parts the first operator in the integrand can
be shifted to the right and converted to an exponential of a multiple
of the adjoint operator so we get
\begin{eqnarray*}
IL & = & \left\langle e^{-\Delta tL^{*}}\left(e^{\Delta tL^{*}}\hat{l(}t)-\left(\Delta t\frac{\partial A_{i}}{\partial x_{i}}+\frac{\left(\Delta t\right)^{2}}{2}L^{*}\left(\frac{\partial A_{i}}{\partial x_{i}}\right)+\ldots\right)-e^{\Delta tT}\hat{l}(t)\right)\right\rangle \\
 & = & \left\langle \hat{l(}t)-e^{-\Delta tL^{*}}\left(\Delta t\frac{\partial A_{i}}{\partial x_{i}}+\frac{\left(\Delta t\right)^{2}}{2}L^{*}\left(\frac{\partial A_{i}}{\partial x_{i}}\right)+\ldots\right)-e^{\Delta t(T-L^{*})}\hat{l}(t)\right\rangle 
\end{eqnarray*}

where the angle bracket is an expectation with respect to the trial
density at time $t$ and on the second line we are using the fact
that $T$ and $L^{*}$ commute due to the autonomous dynamical system
assumption. Expanding out the remaining exponential operators as Taylor
series and retaining only terms to second order:
\begin{eqnarray*}
IL & = & \left\langle \hat{l(}t)-\Delta t\frac{\partial A_{i}}{\partial x_{i}}-\frac{\left(\Delta t\right)^{2}}{2}L^{*}\left(\frac{\partial A_{i}}{\partial x_{i}}\right)+\left(\Delta t\right)^{2}L^{*}\left(\frac{\partial A_{i}}{\partial x_{i}}\right)\right.\\
 &  & \left.-\hat{l}(t)-\Delta t(T-L^{*})\hat{l}(t)-\frac{\left(\Delta t\right)^{2}}{2}(T-L^{*})^{2}\hat{l(}t)\right\rangle \\
 & = & \left\langle -\Delta t\frac{\partial A_{i}}{\partial x_{i}}+\frac{\left(\Delta t\right)^{2}}{2}L^{*}\left(\frac{\partial A_{i}}{\partial x_{i}}\right)+\Delta t\frac{\partial A_{i}}{\partial x_{i}}+\frac{\left(\Delta t\right)^{2}}{2}\left(R^{2}+(T-L^{*})\frac{\partial A_{i}}{\partial x_{i}}\right)\right\rangle \\
 & = & \frac{\left(\Delta t\right)^{2}}{2}\left\langle R^{2}\right\rangle +O(\left(\Delta t\right)^{3})
\end{eqnarray*}

\end{widetext}

where on the second line we are using equations (\ref{Rdef}) and
(\ref{Expand}) as well as the autonomous assumption. There is a remarkable
number of cancellations and a very simple result which nicely extends
the Hamiltonian case. Recall that $R$ vanishes for time evolution
according to the generalized Liouville equation (\ref{Liouville}).

\section{Path integral formalism}

\subsection{The Lagrangian}

Since $IL$ has a straightforward information theoretic interpretation
we set therefore 
\begin{equation}
\mathscr{L}=\frac{1}{2}\left(\Delta t\right)^{2}\left\langle R^{2}\right\rangle _{\hat{\varrho}(t)}\label{positivelagrangian}
\end{equation}

and by analogy with the Onsager and Machlup approach, identify the
action $S$ for this path 
\[
S=\frac{\Delta t}{2}\int\mathscr{L}dt
\]

as the analog of the entropy for our proposed path Boltzmann principle.
To make further progress we need to specify more concretely the trial
density. A convenient choice here is to use maximum entropy theory
with respect to an equilibrium density obtaining
\begin{equation}
\hat{\varrho}\left(\lambda,x\right)=Z^{-1}\left(\lambda\right)\exp\left(\lambda^{t}Q(x)-\beta\psi(x)\right)\label{trialdensity}
\end{equation}

where the function $\psi$ is the analog of invariant quantities for
a Hamiltonian system (often energy in that case). It may only be known
approximately as a quadratic function and need not be assumed neccessarily
to be an invariant of the dynamical system. 

The vector $Q$ is an appropriately chosen set of slow variables for
the system whose moments interest us. For practical reasons they are
usually chosen to be quadratic functions since calculation of expectations
with respect to a trial density often requires a Gaussian density
for tractability. In the case of a forced dissipative turbulent system
a non-trivial equilibrium density occurs and this may be used empirically
(via a direct numerical simulation) to derive $\psi(x)$. In the case
of decaying turbulence when the forcing is absent then one expects
the asymptotic state to be one of no motion meaning that the Lagrange
multipliers $\lambda$ associated with quadratic variables might be
expected to diverge as time proceeds. In such a case it will likely
be expedient to simply set $\beta=0$. Further discussion of these
issues may be found below. 

With the choice of trial density (\ref{trialdensity}), the Lagrangian
is easily evaluated up to an irrelevant additive constant as
\begin{eqnarray}
\mathscr{L} & = & \frac{\left(\Delta t\right)^{2}}{2}\left(\dot{\lambda}^{t}g\dot{\lambda}-2\dot{\lambda}^{t}M+\phi+2\dot{\lambda}^{t}X-2\lambda^{t}Y\right)\label{explicitlagrangian}\\
g_{ij} & \equiv & \left\langle Q_{i},Q_{j}\right\rangle \nonumber \\
M_{i} & \equiv & \left\langle L^{*}Q_{i}\right\rangle \nonumber \\
\phi & \equiv & \lambda^{i}\left\langle L^{*}Q_{i}L^{*}Q_{j}\right\rangle \lambda^{j}\nonumber \\
X_{i} & \equiv & \left\langle \left(Q_{i}-\left\langle Q_{i}\right\rangle \right)\Gamma\right\rangle \nonumber \\
Y_{i} & \equiv & \left\langle \left(L^{*}Q_{i}\right)\Gamma\right\rangle \nonumber \\
\Gamma & \equiv & \frac{\partial A_{i}}{\partial x_{i}}-\beta L^{*}\psi\nonumber 
\end{eqnarray}
The non-negative definite matrix $g(\lambda)$ is the Fisher information
matrix for the random vector $Q$ whose trial densities are the given
maximum entropy statistical model family. Thus it is also the metric
tensor of the trial density manifold. Note also that the scalar function
$\Gamma$ here will vanish identically if $\psi$ happens to be chosen
an invariant function for the dynamical system. In the case it is
a quadratic approximation to such an invariant then $\Gamma$ will
be expected to be small and hence the linear terms in the Lagrangian
will be also. Finally it is worth emphasizing that due to (\ref{positivelagrangian}),
the action is always bounded below and so the Euclidean path integral
we have proposed here is well defined despite the complexity of equation
(\ref{explicitlagrangian}). 

It is clear now that our first and second requirements for a path
measure from section 2 above have been met since all these fields
are specified as functions of $\lambda$ (and hence the moments of
the slow variables via a Legendre transformation) and a Lagrangian
defined based on a clear information theoretic formulation which generalizes
entropy from the equilibrium Boltzmann principle. The third requirement
holds in a certain sense when the system is Hamiltonian but the more
general autonomous case considered here requires further investigation.

\subsection{Paths and endpoints}

We have seen how to associate a weight to a path using the information
loss action. It is clear however that we further require a non-negative
weight for any trial density at a prescribed endpoint time in order
to evaluate it's significance as an approximation for the true probability
density at that time. The path integral proposed by Onsager and Machlup
and generalized here, clearly achieves this objective in a natural
manner (see equation (\ref{pathintegral})). The trial density weight
at a fixed time is termed a \textbf{consistency distribution} and
given it's formulation as a path integral is analogous to a quantum
wavefunction. Indeed it satisfies a (Wick rotated) Schrödinger equation
(see \citep{Kle14}). 

It is important to emphasize the significance of the path integral
here. It is the sum of all path weights with fixed endpoints that
is used to define the consistency distribution. Such a path integral
is \textbf{not} \textbf{in general} a monotonic function of the extremal
action which applies to a path satisfying the Euler-Lagrange equations
for $\mathscr{L}$. Such an action is termed the classical action
$S_{cl}$ in the path integral literature. In the special case that
the Lagrangian $\mathscr{L}$ is a \textbf{quadratic} function of
$\lambda$ it may however be proven (see e.g. \citep{schulman2012techniques}
Chapter 6) that 
\[
\int D\lambda W\left[\lambda\right]=B\exp\left(-S_{cl}\right)
\]

with $B$ not dependent on the value of $\lambda$ at the path endpoint.
Clearly in that simple case (which also happens to be the original
Onsager-Machlup one) only the extremal actions are important in defining
the consistency distribution. In most realistic situations however
$\mathscr{L}$ is not quadratic and so this simplifying situation
may not apply.

More concretely, one can imagine a scenario when the extremal paths
between two sets of endpoints have the same minimal action but the
path integral is quite different. This situation is sketched in Figure
1. 
\begin{figure}[h]
\begin{centering}
\includegraphics[scale=0.33]{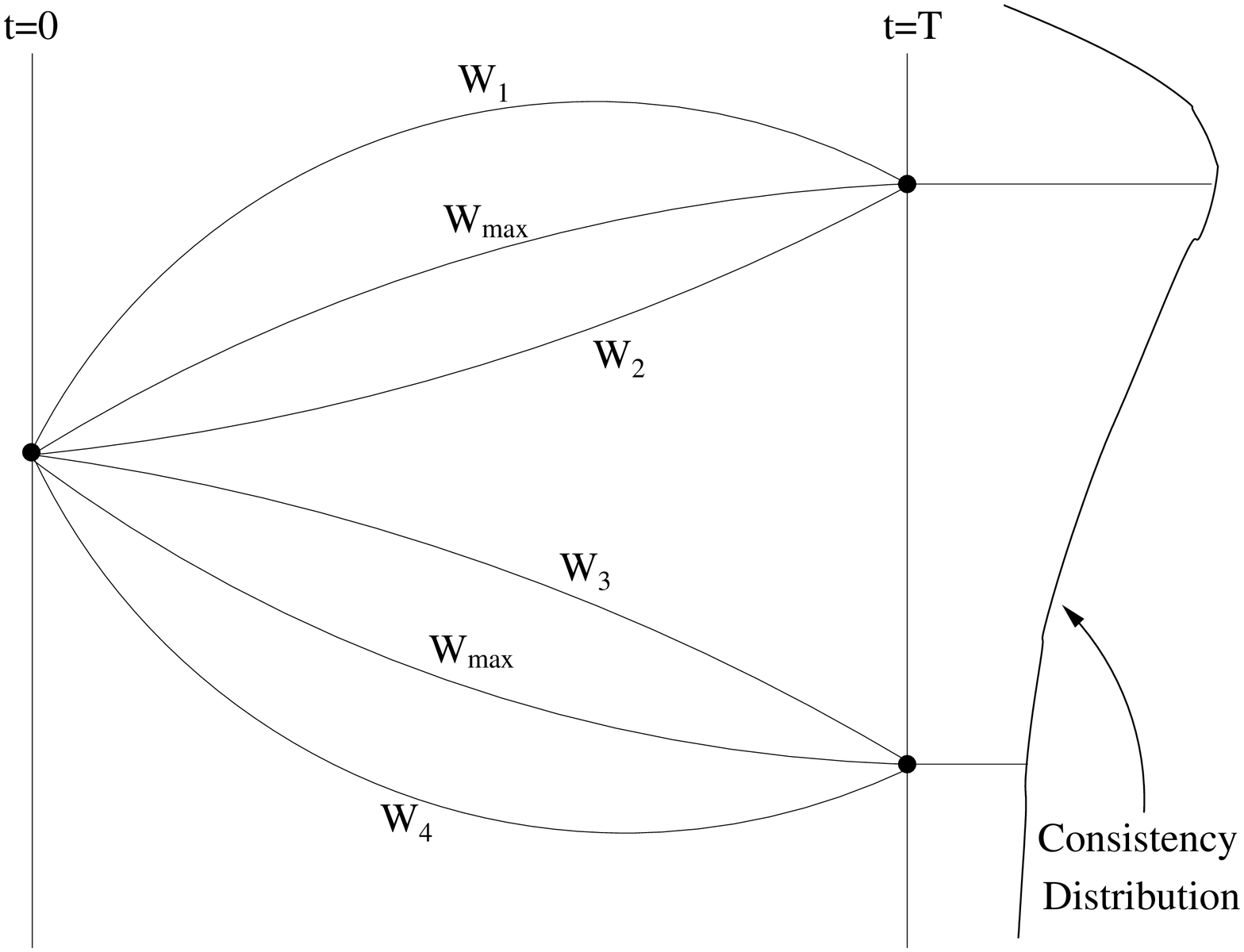}
\par\end{centering}

\caption{The importance of a path integral versus extremal actions in defining
the consistency distribution. In the example sketch we have the situation
that $W_{3},W_{4}<<W_{1},W_{2}<W_{max}$ which is when ``quantum''
effects are significant.}
\end{figure}

\section{Discussion}

The present contribution is intended as a documentation of a closure
formalism which works \textbf{in principle }for a wide class of realistic
dynamical systems including those pertaining to turbulence. We discuss
now the practical problems involved in implementing this formalism. 

One issue to consider is the choice of approximating or trial densities.
In section 4 we posited that some kind of exponential family of densities
was appropriate and this assumption allowed the \textbf{formal} completion
of the calculation of the Lagrangian. A key consideration here is
whether a member of the chosen trial density manifold could be considered
to be a reasonable approximation to the non-equilibrium densities
of the full system. If this is not so one should not expect a priori
the present closure to perform well. This is an important issue since
a complete calculation of the Lagrangian involves knowledge of the
trial density partition function $Z$ as well as its various moments.
In the case that a Gaussian trial density is a good approximating
density then this calculation is straightforward. It is clear however
that a more general class of trial densities may need to be chosen
carefully in order to balance approximating accuracy against the ability
to analytically evaluate moments and the partition function. 

It is also important to consider the statistical behaviour of the
unresolved or fast variables within the system. The implicit assumption
of the present closure method is that on the slow timescale of interest,
the fast variable statistics of the system are effectively relaxed
to those applicable for an equilibrium density. Referring to the ansatz
trial density (\ref{trialdensity}) the only dependency of the density
on fast variables is assumed to come through the function $\psi(x)$.
For a forced dissipative turbulent system the obvious choice for this
function is one such that $C\exp(-\beta\psi(x))$ is close to the
equilibrium density of the system since that ensures the trial densities
have the correct statistical properties for the fast variables (at
least on the timescale of interest). Knowledge of such a density may
be in general difficult to obtain theoretically and instead require
a direct numerical simulation of the system. For a decaying turbulent
system, the asymptotic state is usually one of no motion in which
case the $\exp(-\beta\psi(x))$ piece could plausibly be replaced
simply by the products of delta functions for the fast variables since
these are the limits of Gaussian densities as the variance go to zero. 

A further issue to consider is the numerical evaluation of the path
integrals proposed here. General Markov Chain Monte Carlo methods
for such integrals are well known (see, for example, \citep{ceperley1995path})
and these have been very recently adapted by the author to the present
situation (see \citep{RK15}). One constraint here is that the number
of coarse graining variables used needs to be manageably small in
order that the numerical calculations remain tractable. A further
constraint concerns the time discretisation used to approximate the
action integral since the numerical cost of the MCMC method depends
significantly on the number of time ordinates used as documented by
Ceperley and co-workers. It should be noted that a potential advantage
of the path integral approach as opposed to other methods used previously
by the author and collaborator (see \citep{kleeman2012nonequilibrium})
is that it is applicable to far from statistical equilibrium situations.
The other method relies on a Taylor expansion of an associated Hamilton-Jacobi
equation about the equilibrium density.


\begin{thebibliography}{15}%
\makeatletter
\providecommand \@ifxundefined [1]{%
 \@ifx{#1\undefined}
}%
\providecommand \@ifnum [1]{%
 \ifnum #1\expandafter \@firstoftwo
 \else \expandafter \@secondoftwo
 \fi
}%
\providecommand \@ifx [1]{%
 \ifx #1\expandafter \@firstoftwo
 \else \expandafter \@secondoftwo
 \fi
}%
\providecommand \natexlab [1]{#1}%
\providecommand \enquote  [1]{``#1''}%
\providecommand \bibnamefont  [1]{#1}%
\providecommand \bibfnamefont [1]{#1}%
\providecommand \citenamefont [1]{#1}%
\providecommand \href@noop [0]{\@secondoftwo}%
\providecommand \href [0]{\begingroup \@sanitize@url \@href}%
\providecommand \@href[1]{\@@startlink{#1}\@@href}%
\providecommand \@@href[1]{\endgroup#1\@@endlink}%
\providecommand \@sanitize@url [0]{\catcode `\\12\catcode `\$12\catcode
  `\&12\catcode `\#12\catcode `\^12\catcode `\_12\catcode `\%12\relax}%
\providecommand \@@startlink[1]{}%
\providecommand \@@endlink[0]{}%
\providecommand \url  [0]{\begingroup\@sanitize@url \@url }%
\providecommand \@url [1]{\endgroup\@href {#1}{\urlprefix }}%
\providecommand \urlprefix  [0]{URL }%
\providecommand \Eprint [0]{\href }%
\providecommand \doibase [0]{http://dx.doi.org/}%
\providecommand \selectlanguage [0]{\@gobble}%
\providecommand \bibinfo  [0]{\@secondoftwo}%
\providecommand \bibfield  [0]{\@secondoftwo}%
\providecommand \translation [1]{[#1]}%
\providecommand \BibitemOpen [0]{}%
\providecommand \bibitemStop [0]{}%
\providecommand \bibitemNoStop [0]{.\EOS\space}%
\providecommand \EOS [0]{\spacefactor3000\relax}%
\providecommand \BibitemShut  [1]{\csname bibitem#1\endcsname}%
\let\auto@bib@innerbib\@empty
\bibitem [{\citenamefont {Kleeman}(2015)}]{Kle14}%
  \BibitemOpen
  \bibfield  {author} {\bibinfo {author} {\bibfnamefont {R.}~\bibnamefont
  {Kleeman}},\ }\href@noop {} {\bibfield  {journal} {\bibinfo  {journal} {J.
  Stat. Phys.}\ }\textbf {\bibinfo {volume} {158}},\ \bibinfo {pages} {1271}
  (\bibinfo {year} {2015})}\BibitemShut {NoStop}%
\bibitem [{\citenamefont {Turkington}(2013)}]{turkington2012optimization}%
  \BibitemOpen
  \bibfield  {author} {\bibinfo {author} {\bibfnamefont {B.}~\bibnamefont
  {Turkington}},\ }\href@noop {} {\bibfield  {journal} {\bibinfo  {journal} {J.
  Stat. Phys}\ }\textbf {\bibinfo {volume} {152}},\ \bibinfo {pages} {569}
  (\bibinfo {year} {2013})}\BibitemShut {NoStop}%
\bibitem [{\citenamefont {Kleeman}\ and\ \citenamefont
  {Turkington}(2014)}]{kleeman2012nonequilibrium}%
  \BibitemOpen
  \bibfield  {author} {\bibinfo {author} {\bibfnamefont {R.}~\bibnamefont
  {Kleeman}}\ and\ \bibinfo {author} {\bibfnamefont {B.~E.}\ \bibnamefont
  {Turkington}},\ }\href@noop {} {\bibfield  {journal} {\bibinfo  {journal}
  {Comm. Pure Appl. Math.}\ }\textbf {\bibinfo {volume} {67}},\ \bibinfo
  {pages} {1905} (\bibinfo {year} {2014})}\BibitemShut {NoStop}%
\bibitem [{\citenamefont {Salmon}(1998)}]{salmon-book}%
  \BibitemOpen
  \bibfield  {author} {\bibinfo {author} {\bibfnamefont {R.}~\bibnamefont
  {Salmon}},\ }\href@noop {} {\emph {\bibinfo {title} {Lectures on Geophysical
  Fluid Dynamics}}}\ (\bibinfo  {publisher} {Oxford Univ. Press, New York},\
  \bibinfo {year} {1998})\BibitemShut {NoStop}%
\bibitem [{Note1()}]{Note1}%
  \BibitemOpen
  \bibinfo {note} {For pedagogical reasons we are considering a finite set of
  dynamical system variables which implies some kind of truncation of the fluid
  system at a fine scale. Generalizations to a continuum of variables are
  straighforward. Note also that the summation convention is being used on
  repeated indices.}\BibitemShut {Stop}%
\bibitem [{Note2()}]{Note2}%
  \BibitemOpen
  \bibinfo {note} {Note that $J$ transforms as a second rank tensor under
  non-singular changes of dynamical variables.}\BibitemShut {Stop}%
\bibitem [{\citenamefont {Onsager}\ and\ \citenamefont
  {Machlup}(1953)}]{onsager1953fluctuations}%
  \BibitemOpen
  \bibfield  {author} {\bibinfo {author} {\bibfnamefont {L.}~\bibnamefont
  {Onsager}}\ and\ \bibinfo {author} {\bibfnamefont {S.}~\bibnamefont
  {Machlup}},\ }\href@noop {} {\bibfield  {journal} {\bibinfo  {journal} {Phys.
  Rev.}\ }\textbf {\bibinfo {volume} {91}},\ \bibinfo {pages} {1505} (\bibinfo
  {year} {1953})}\BibitemShut {NoStop}%
\bibitem [{\citenamefont {Risken}(1989)}]{risk89}%
  \BibitemOpen
  \bibfield  {author} {\bibinfo {author} {\bibfnamefont {H.}~\bibnamefont
  {Risken}},\ }\href@noop {} {\emph {\bibinfo {title} {The Fokker-Plank
  Equation}}},\ \bibinfo {edition} {2nd}\ ed.\ (\bibinfo  {publisher} {Springer
  Verlag, Berlin},\ \bibinfo {year} {1989})\BibitemShut {NoStop}%
\bibitem [{\citenamefont {Oettinger}(2005)}]{oett}%
  \BibitemOpen
  \bibfield  {author} {\bibinfo {author} {\bibfnamefont {H.~C.}\ \bibnamefont
  {Oettinger}},\ }\href@noop {} {\emph {\bibinfo {title} {Beyond equilibrium
  thermodynamics}}}\ (\bibinfo  {publisher} {Wiley-Interscience},\ \bibinfo
  {address} {Hoboken, New Jersey},\ \bibinfo {year} {2005})\BibitemShut
  {NoStop}%
\bibitem [{\citenamefont {Zubarev}(1974)}]{Zub74}%
  \BibitemOpen
  \bibfield  {author} {\bibinfo {author} {\bibfnamefont {D.~N.}\ \bibnamefont
  {Zubarev}},\ }\href@noop {} {\emph {\bibinfo {title} {Nonequilibrium
  {S}tatistical {T}hermodynamics}}}\ (\bibinfo  {publisher} {Plenum Press, New
  York},\ \bibinfo {year} {1974})\BibitemShut {NoStop}%
\bibitem [{\citenamefont {Amari}\ and\ \citenamefont {Nagaoka}(2000)}]{ama00}%
  \BibitemOpen
  \bibfield  {author} {\bibinfo {author} {\bibfnamefont {S.}~\bibnamefont
  {Amari}}\ and\ \bibinfo {author} {\bibfnamefont {H.}~\bibnamefont
  {Nagaoka}},\ }\href@noop {} {\emph {\bibinfo {title} {Methods of Information
  Geometry}}}\ (\bibinfo  {publisher} {Translations of Mathematical Monographs,
  AMS, Oxford University Press},\ \bibinfo {year} {2000})\BibitemShut {NoStop}%
\bibitem [{Note3()}]{Note3}%
  \BibitemOpen
  \bibinfo {note} {We refer to this equation as a \protect \uline {generalized}
  Liouville equation as opposed to the classical such equation of (\ref
  {classical}).}\BibitemShut {Stop}%
\bibitem [{\citenamefont {Schulman}(2005)}]{schulman2012techniques}%
  \BibitemOpen
  \bibfield  {author} {\bibinfo {author} {\bibfnamefont {L.~S.}\ \bibnamefont
  {Schulman}},\ }\href@noop {} {\emph {\bibinfo {title} {Techniques and
  applications of path integration}}}\ (\bibinfo  {publisher} {Dover},\
  \bibinfo {year} {2005})\BibitemShut {NoStop}%
\bibitem [{\citenamefont {Ceperley}(1995)}]{ceperley1995path}%
  \BibitemOpen
  \bibfield  {author} {\bibinfo {author} {\bibfnamefont {D.~M.}\ \bibnamefont
  {Ceperley}},\ }\href@noop {} {\bibfield  {journal} {\bibinfo  {journal} {Rev
  Mod Phys}\ }\textbf {\bibinfo {volume} {67}},\ \bibinfo {pages} {279}
  (\bibinfo {year} {1995})}\BibitemShut {NoStop}%
\bibitem [{\citenamefont {{R. Kleeman}}(2015)}]{RK15}%
  \BibitemOpen
  \bibfield  {author} {\bibinfo {author} {\bibnamefont {{R. Kleeman}}},\
  }\href@noop {} {\bibfield  {journal} {\bibinfo  {journal} {J. Stat. Phys.}\ }
  (\bibinfo {year} {2015})},\ \bibinfo {note} {in preparation}\BibitemShut
  {NoStop}%
\end{thebibliography}
\end{document}